\documentclass[aps,prb,twocolumn,superscriptaddress]{revtex4-1}
\usepackage[compat=1.1.0]{tikz-feynman}
\usepackage{graphicx,amssymb,amsmath,bm,color,bbm}
\newcommand{\javi}[1]{\textcolor{black}{#1}}%!!!!!COMMENT THE MACRO ABOVE OUT AND UNCOMMENT THIS ONE!!!!

\newcommand{\jav}[1]{\textcolor{black}{#1}}

\begin{document}

\title{Collective modes for helical edge state interacting with quantum light}
\author{Bal\'azs Gul\'acsi}
\email{gulacsi@phy.bme.hu}
\affiliation{Department of Theoretical Physics and MTA-BME Lend\"ulet Topology and Correlation Research Group,Budapest University of Technology and Economics, 1521 Budapest, Hungary}
\author{Bal\'azs D\'ora}
\affiliation{Department of Theoretical Physics and MTA-BME Lend\"ulet Topology and Correlation Research Group,Budapest University of Technology and Economics, 1521 Budapest, Hungary}
\date{\today}

\begin{abstract}
We investigate the light-matter interaction between the edge state of a 2D topological insulator and quantum electromagnetic field.  
The interaction originates from the Zeeman term between the spin of the edge electrons and the magnetic field, and also through the Peierls substitution. 
The continuous U(1) symmetry of the system in the absence of the vector potential reduces into discrete time reversal symmetry in the presence of the vector potential.
Due to light-matter interaction, a superradiant ground state emerges with spontaneously broken time reversal symmetry, accompanied by a net photocurrent along the edge, generated by the vector potential of the quantum light. 
The spectral function of the photon field reveals polariton continuum excitations above a threshold energy, corresponding to a Higgs mode and another low energy collective mode due to 
the phase fluctuations of the ground state. This collective mode is a zero energy Goldstone mode that arises from the 
broken continuous U(1) symmetry in the absence of the vector potential, and acquires  finite a gap in the presence of the vector potential. 
The optical conductivity of the edge electrons is calculated using the random phase approximation by taking the fluctuation of the order parameter into account. 
It contains the collective modes as a Drude peak with renormalized effective mass, which moves to finite frequencies as the symmetry of 
the system is lowered by the inclusion of the vector potential. 
\end{abstract}

% insert suggested PACS numbers in braces on next line
% insert suggested keywords - APS authors don't need to do this
%\keywords{}

%\maketitle must follow title, authors, abstract, \pacs, and \keywords
\maketitle

% body of paper here - Use proper section commands
% References should be done using the \cite, \ref, and \label commands
\section{Introduction}
Interaction between light and matter are the basis of wide range modern technologies, including lasers, LEDs and computers. From a theoretical point of 
view, even the simplest quantum optical models describing light-matter interaction, like the Dicke model\cite{dicke}, offer a variety of interesting phenomena 
such as quantum phase transitions and quantum chaos\cite{emarybrandes}. In the Dicke model a single mode of electromagnetic field interacts with an ensemble of two level atoms. 
The ground state of such a system is composed of unexcited atoms and an unpopulated photon mode at weak coupling. However at a critical coupling strength the atoms are collectively 
excited and the photon mode becomes macroscopically populated, coined  superradiance.  
The recent realizations of this phase transition has opened a way to studying other relating phenomena\cite{baumann,exp1,exp2} in the controlled environment of cold atomic physics.

%Ever since its prediction and experimental discovery\cite{hasankane} topological insulators have generated a great amount of excitement in the physics community. In this state of matter, conducting edge states form on the surface of topological insulators due to the topologically non trivial nature of its bulk insulating band structure.\cite{bulk} These topologically protected edge states can serve as building blocks of upgrading conventional computer physical memory, a variety of spintronics devices and most of all realising practical quantum computers.\cite{qcomp} 

Subjecting quantum gases to cavity modes can produce remarkable changes in both the atomic gas and the cavity field. 
For instance, a driven Bose--Einstein condensate placed in a cavity undergoes a quantum phase transition that corresponds to the self-organization of atoms from homogeneous into a
periodically patterned distribution above a critical driving strength and the cavity field acquires a nonzero expectation value\cite{nagy,domokos,piazza2,bosee1}. 
Due to cavity-induced long-range interactions between atoms the Bose--Hubbard model inside a cavity exhibits a rich phase diagram, the interacting bosons transition 
from a normal phase to a superfluid phase and at even stronger pumping a self-organized Mott insulator phase\cite{baki,landig}. Many different proposals have been put 
forward to realize the self-organization of more complex quantum phases reaching from the Mott-insulator and disordered structures to phases with spin-orbit 
coupling\cite{boseglass,zhangdicke,cikk1,cikk2}. Fermionic quantum gases inside a cavity can also exhibit superradiant phenomena and can self-organize into 
topologically non-trivial phases\cite{fermioncav}. The superradiant light generation in the transversely driven cavity mode induces a cavity-assisted spin-orbit 
coupling and opens a bulk gap at half filling for a degenerate Fermi gas in a cavity. This mechanism can simultaneously drive a topological phase transition in the system, yielding a topological superradiant state\cite{topferm,trif,piazza1}. 

In a topological phase, matter possesses exceptional properties such as edge or surface states that are protected from small external perturbations\cite{hasankane,bulk}. 
These protected edge states of topological insulators (TI) can serve as building blocks of upgrading conventional computer physical memory, a variety of spintronics devices and most of all realising practical quantum computers\cite{qcomp}. 

In the present work we are combining TIs with cavity physics and investigate the interaction between a spin polarized edge state of a quantum spin Hall insulator with linear dispersion and a single mode of circularly polarized quantum 
electromagnetic field inside a cavity. The spin Hall insulator can be realized using either condensed matter\cite{hasankane} or cold atomic setting\cite{goldman}. 
The coupling between a condensed matter realized topological insulator edge state and quantum light field includes the Zeeman term and Peierls substitution.
 However, in ultracold bose and fermi gases, the charge neutrality of the atoms requires to engineer artificial vector potentials, 
which act similarly to magnetic fields for charged particles\cite{synt,artmag1,artmag2,artmag3}. 
A single photon mode with fixed helicity can be realized by selection from a ladder of cavity modes by placing a dispersive element into the cavity such as a prism or nonlinear dielectric material. 
The system might also be implemented using circuit quantum electrodynamical systems\cite{cqed}. 

The structure of this paper is as follows: in section~\ref{kato}, we introduce the Hamiltonian of our system, 
illustrate its properties and then use mean field theory to determine its ground state. In section~\ref{Pf}, we focus on the photon field, calculate its spectral function by taking Gaussian fluctuations into account on top of the mean field solutions and 
discuss its properties. In the last section we investigate the frequency dependent optical conductivity along the edge to reveal the subtle effect of light-matter interaction on electronic transport.

\section{\label{kato}The model}
% Put \label in argument of \section for cross-referencing
%\section{\label{}}
%\subsection{}
%\subsubsection{}
Our system consists of \jav{spin-momentum locked edge electrons} of a quantum spin Hall
insulator with linear \jav{momentum} and a single mode of circularly polarized quantum electromagnetic field of a cavity. 
\jav{Treating the cavity field as having its own quantum dynamics enables us to describe the system in equilibrium and such the use of concepts like the existence of a ground state are justified\cite{trif}.}

The light-matter interaction originates from the Zeeman term between the edge spins and magnetic field and from another term through the Peierls substitution. The full Hamiltonian of the \jav{system} is
\begin{gather}
H=\omega a^\dagger a + \sum_p 2vpS_p^z+\frac{g_A}{\sqrt L}\sum_p \left(a+a^\dagger\right)S_p^z\nonumber \\ +\frac{g}{\sqrt L}\sum_p \left(aS_p^++a^\dagger S_p^-\right)\label{HAM},
\end{gather}
where the first term is the energy of the cavity mode: $\omega$ being the photon frequency, $a^\dagger$ creates a photon with positive helicity. The second term of Eq.~\eqref{HAM} describes the spin polarized edge \jav{electrons} with $S_p^z=\frac{1}{2}(c^\dagger_{p\uparrow}c_{p\uparrow}-c^\dagger_{p\downarrow}c_{p\downarrow})$, where $c^\dagger_{p\sigma}$ creates an edge electron with momentum $p$ and spin $\sigma=(\uparrow,\downarrow)$, $v$ is the Fermi velocity $(\hbar=1)$. Since the edge Hamiltonian is linear in momentum the \jav{the electromagnetic field's vector potential appears due to the Peierls substitution which} is characterized by the third term with $g_A=\frac{ev}{\sqrt{\omega\varepsilon_0}}=\sqrt{\frac{\tilde g_A}{\omega}}$ the coupling strength of this interaction and $L$ being the dimensionless length of
the edge, which is defined as the number of edge sites times the lattice constant which is taken to be unity. \jav{The number of electrons that occupy the edge state and interact with the quantum light is therefore proportional to $L$.}
The last one is the Zeeman term with $g=\frac{g_e\mu_B}{2c}\sqrt{\frac{\omega}{\epsilon_0}}=\sqrt{\tilde g\omega}$, where $g_e$ is the effective g-factor of the edge electrons, 
$\mu_B$ is the Bohr magneton, $c$ the speed of light, $\epsilon_0$ is the vacuum permittivity and finally, $S_p^+=~c^\dagger_{p\uparrow}c_{p\downarrow},S_p^-=~c^\dagger_{p\downarrow}c_{p\uparrow}$. 
\jav{A detailed derivation of the Hamiltonian is done in the Appendix.} We assume that the Zeeman coupling is always stronger than the vector potential interaction: $g_A<g$, which is satisfied if the photon frequency is $\omega>m_{eff}cv$, with $m_{eff}$ effective edge 
electron mass. \jav{The topological insulator that supports our linear edge state must have a band gap $W$, and throughout the calculations we assume the energies to be much smaller than this band gap so the effects of the insulator's bulk states can be neglected.} It is important to remark, that the absence of a counter rotating term\cite{emarybrandes} \jav{in Eq.~\eqref{HAM}} is the result of the electromagnetic field being circularly polarized. Furthermore, the wave 
vector of the cavity mode is assumed to be perpendicular to the direction of the topological insulator's edge state.
A term identical to the vector potential term can also be generated if the propagation direction of the quantum light has an angle of incidence $\theta$ with the edge state. 
The coupling strength of this term is then $g\sin\theta$ and the coupling strength of the Zeeman term becomes $g\cos\theta$.

\begin{figure}[t]
\includegraphics[width=8cm,height=4.5cm]{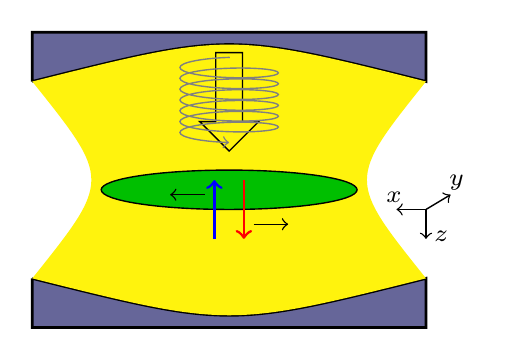}% Here is how to import EPS art
\caption{\label{rajz} \jav{Schematic illustration of a quantum spin Hall insulator with spin filtered edge state placed inside an optical cavity. The green ellipse-like object represents the quantum spin Hall insulator that supports the edge states, while the blue and red arrows the spin-momentum locked electrons on the edge. The wave vector of the cavity mode is perpendicular to the direction of the edge electrons.}}
\end{figure}

Let us first discuss the case when $g_A=0$, which makes Eq.~\eqref{HAM} an inhomogeneous Dicke model\cite{inhom}. Without $g_A$ the Hamiltonian exhibits U(1) symmetry, indeed $e^{i\phi(a^\dagger a+\sum_p S_p^z)}$ leaves the Hamiltonian invariant and the total number of excitations $N=a^\dagger a+\sum_p S_p^z$ is a constant of motion. Time reversal is the other symmetry of the system without $g_A$, using $\mathcal T=e^{i\pi(a^\dagger a+\sum_p S_p^y)}\mathcal K$, where $\mathcal K$ is complex conjugation, the Hamiltonian is unchanged:
\begin{eqnarray*}
\mathcal T (a,a^\dagger,p,S_p^\pm , S_p^z)\mathcal T^{-1}=(-a,-a^\dagger,-p,-S_p^\pm,-S_p^z).
\end{eqnarray*}

Reintroducing $g_A$ destroys the U(1) symmetry which means that the total number of excitations are no longer conserved. Since $\mathcal T$ leaves the vector potential term invariant, the sole symmetry of the \jav{full} system is time reversal. Furthermore, the system is integrable when U(1) symmetry is present and its mean field solution coincides with the exact solution\cite{eastlittle}. One can argue that if we integrate out the photon degree of freedom, the resulting effective electron-electron interaction has the form $-\tilde g L^{-1}\sum_{p,p'}S^+_p S^-_{p'} $ which describes an infinite range and constant strength interaction that makes the mean field results in the thermodynamic limit ($L\to\infty$) exact. The same argument holds when $g_A\neq0$. After integrating out the photon field it yields an effective interaction:
\begin{gather}
H_{eff}=-\frac{\tilde g}{L}\sum_{p,p'}S^+_p S^-_{p'}\nonumber \\ -\frac{\sqrt{\tilde g\tilde{g_A}}}{\omega L}\sum_{p,p'}\left(S^+_p S^z_{p'}+S^z_p S^-_{p'}\right)-\frac{\tilde{g_A}}{\omega^3L}\sum_{p,p'}S_p^zS_{p'}^z\label{heffga},
\end{gather}
which also describes infinite range and constant strength interactions between electrons, therefore the mean field solution in the thermodynamic limit is \jav{still} exact. The last term in Eq.~\eqref{heffga} is a ferromagnetic coupling between electron spins mediated by the vector potential of the cavity field, as we will see this results in a generated photocurrent along the edges. 

\subsection{Mean field theory}

In the thermodynamic limit the photon field becomes macroscopically occupied\cite{emaryprl}: $\langle a\rangle=\sqrt{nL}e^{i\varphi}$, the system is in a superradiant phase. The mean field \jav{description} means that we replace \jav{the bosonic operators with their mean value and then the} Hamiltionian \jav{in Eq.~\eqref{HAM}} becomes:
\begin{eqnarray}
\label{MF}
H_{MF}=\omega nL+\sum_p\begin{pmatrix} c^\dagger_{p\uparrow}&c^\dagger_{p\downarrow}\end{pmatrix}\begin{pmatrix} \varepsilon_p & \Delta\\ \Delta^* &-\varepsilon_p \end{pmatrix}\begin{pmatrix} c_{p\uparrow}\\c_{p\downarrow}\end{pmatrix},
\end{eqnarray}
where $\varepsilon_p=vp+g_A\sqrt n\cos\varphi$ and $\Delta=g\sqrt n e^{i\varphi}$. Eq.~\eqref{MF} is easily diagonalized by the Bogoliubov transformation:
\begin{eqnarray}
\begin{pmatrix} c_{p\uparrow}\\c_{p\downarrow}\end{pmatrix}=\begin{pmatrix} \cos\vartheta & e^{i\varphi}\sin\vartheta\\ -e^{-i\varphi}\sin\vartheta &\cos\vartheta \end{pmatrix}\begin{pmatrix} d_{p+}\\d_{p-}\end{pmatrix},
\label{bog}
\end{eqnarray}
where $\tan 2\vartheta=-|\Delta|/\varepsilon_p$ and Eq.~\eqref{MF} becomes:
\begin{eqnarray}
H_{MF}=\omega nL+\sum_{p,\alpha}E_\alpha(p) d^\dagger_{p\alpha}d_{p\alpha}.
\label{HMF}
\end{eqnarray}
Here, $\alpha=\pm 1$ and $E_\alpha(p)=\alpha E_p$ with: %the fermionic spectrum aquires a gap $|\Delta|=~g\sqrt n$ and becomes shifted:
\begin{eqnarray}
\label{Ep}
E_p=\sqrt{(vp+g_A\sqrt n\cos\varphi)^2+g^2n}.
\end{eqnarray}

\begin{figure}[t]
\includegraphics[width=9cm,height=4.5cm]{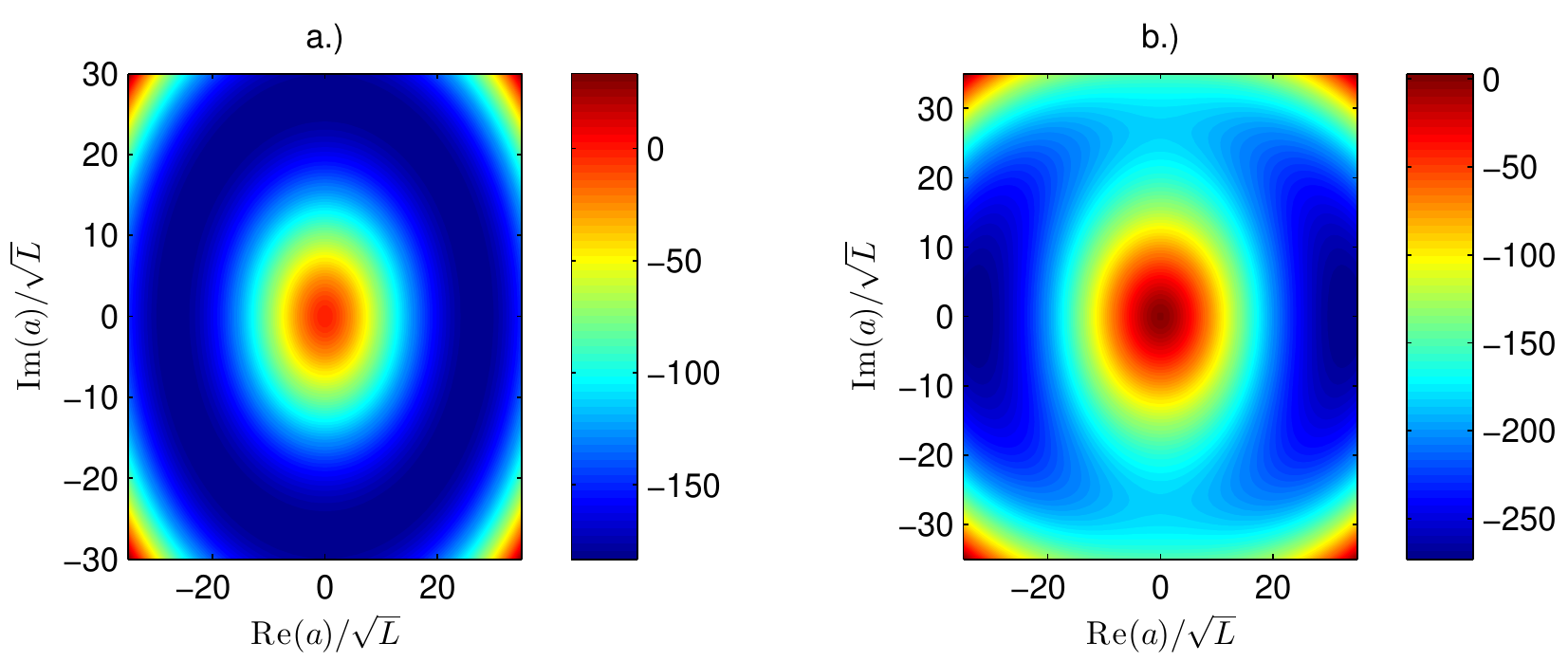}% Here is how to import EPS art
\caption{\label{ener} The contourplot for the total ground-state energy. In panel a.) $g_A=0$, the energy exhibits a mexican hat structure with $\varphi\in[0,2\pi]$ one can sweep through the ground-state manifold with no energy cost. In panel b.) $\rho g_A=0.1$ the minimum bends toward the real axis when $\varphi=0$ and $\pi$, so tunneling between the two degenerate ground-state will require a finite amount of energy. The other parameters used: $\rho W=100$, $\rho\omega=1$, $\rho g=1$.}
\end{figure}

The mean field parameters $(n,\varphi)$ which are understood as the mean photon number \jav{density} and the phase of the order parameter, respectively, can be calculated by minimizing the total ground-state energy $(E_{gs})$. At half filling the $\alpha=-1$ band is fully populated and the ground-state energy with $W$ cutoff energy and $\rho=1/v\pi$ 1D density of states is:

\begin{gather}
\label{en}
\frac{E_{gs}}{L}=\omega n -\frac{1}{L}\sum_p\sqrt{\varepsilon_p^2+|\Delta|^2}=\\-\frac{\rho W^2}{2}+\left(\omega-\frac{\rho g_A^2}{2}\cos^2\varphi\right)n-\frac{\rho |\Delta|^2}{4}\left(1+\ln\frac{4W^2}{|\Delta|^2}\right)\nonumber
\end{gather}
With $g_A=0$ the energy exhibits a mexican hat structure in the Re$\langle a\rangle-$Im$\langle a\rangle$ space, see Fig.~\ref{ener}., the phase remains undetermined and the ground-state is infinitely degenerate due to U(1) symmetry. When $g_A\neq0$ the mexican hat structure developes two minima along Re$\langle a\rangle$ and the minimum energy appears when $\cos^2\varphi=1$. The ground-state is now doubly degenerate due to time reversal symmetry which is spontaneously broken in the emerging superradiant phase.

\begin{figure}[h]
\includegraphics[width=6.5cm,height=5cm]{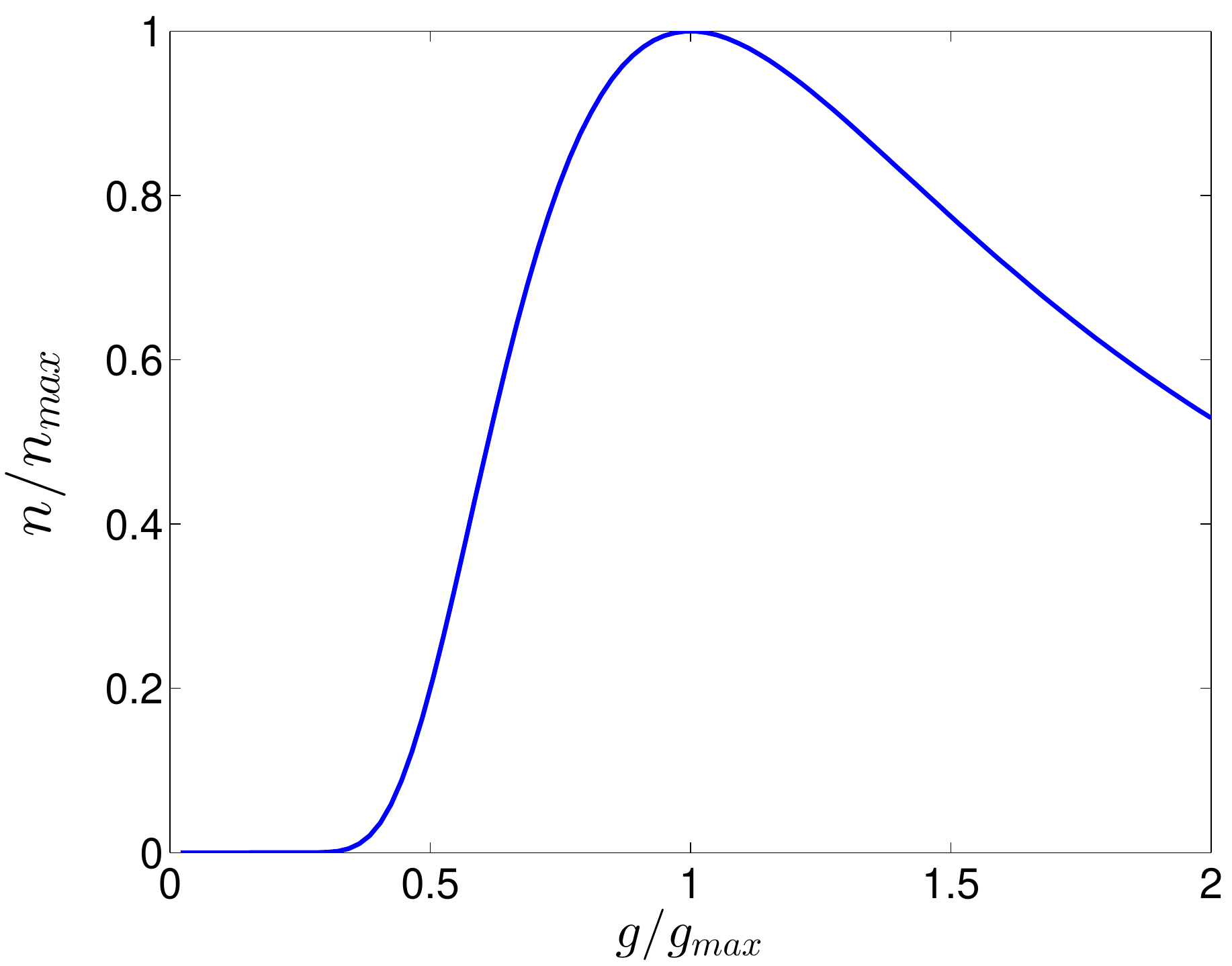}% Here is how to import EPS art
\caption{\label{fotnum} The mean photon number \jav{density} as the function of the Zeeman coupling. The parameters used: $\rho\omega=1$, $\rho g_A=0.1$.}
\end{figure}

By carrying out the minimalization of Eq.~\eqref{en} we find the phase and mean photon number \jav{density} to be:
\begin{eqnarray}
\varphi = m\pi,\quad m\in \mathbb Z\quad\quad\quad\nonumber\\
n=\frac{4W^2}{g^2}\exp\left(\frac{2\rho g_A^2-4\omega}{\rho g^2}\right).
\label{mfp}
\end{eqnarray}
The mean photon number \jav{density} as the function of the Zeeman coupling $g$ is always strictly positive: $n(g)>0$. Since we assume the vector potential coupling strength is always smaller than the Zeeman coupling ($g> g_A$), the photon number \jav{density} has a maxima at $\rho g_{max}=\sqrt{4\rho\omega-2\rho^2g_A^2}$ with $n_{max}=2\rho W^2\exp(-1)/(2\omega-\rho g_A^2)$\jav{, see Fig.~\ref{fotnum}. Detecting the photon number can be achieved by various quantum nondemolition measurements\cite{fdet1,fdet2}, for example subjecting the field to a quasiresonant beam of Rydberg atoms and measuring the resulting phase shift of the atomic wave function\cite{fotondet}.}

The conventional Dicke model of two level atoms predicts a phase transition at a critical coupling constant\cite{emarybrandes}. On the other hand, considering the effects of the vector potential as a diamagnetic term, it can be shown that the condition for a stable superradiant phase is never satisfied due to the Thomas--Reiche--Kuhn sum rule for atomic systems\jav{\cite{nogo,unstab,polini}.} This is known as a no-go theorem. The main differences between our model and the conventional Dicke model is that the edge electrons have linear dispersion while the Dicke two level atoms have a constant energy difference between levels and the fact that here the vector potential appears as a linear term while the diamagnetic term is quadratic. \jav{The Dicke critical coupling constant is proportional to the square root of the energy difference between the atomic levels, for linearly dispersive two level systems this critical value reduces to zero.} 
Eq.~\eqref{mfp} shows us that for arbitrary small $g>0$ and for every $g_A$ the photon states are macroscopically occupied\jav{: $\langle a^\dagger a\rangle=nL$, thus our system is always in its superradiant phase as nothing prevents the phase transition from occuring.}

\subsubsection{Properties of the ground-state}

In the emerging superradiant ground-state time reversal symmetry is spontaneously broken. This fact is proven by the magnetic properties of this state. Indeed using the Bogoliubov transformation, we get the spin expectation values as
\begin{gather}
\langle S^x\rangle=\frac{1}{L}\sum_p\langle S_p^x\rangle=-\frac{1}{2L}\sum_p\frac{|\Delta|\cos\varphi}{\sqrt{\varepsilon_p^2+|\Delta|^2}},\nonumber\\
\langle S^y\rangle=\frac{1}{L}\sum_p\langle S_p^y\rangle=\frac{1}{2L}\sum_p\frac{|\Delta|\sin\varphi}{\sqrt{\varepsilon_p^2+|\Delta|^2}},\\
\langle S^z\rangle=\frac{1}{L}\sum_p\langle S_p^z\rangle=-\frac{1}{2L}\sum_p\frac{\varepsilon_p}{\sqrt{\varepsilon_p^2+|\Delta|^2}}.\nonumber
\end{gather} 
The magnetization along the $x$ and $z$ axis are nonzero and their measured value would determine $\varphi$. The calculations show us that the magnetization is proportional to the gap $|\Delta|$:
\begin{gather}
\langle S^x\rangle=-\frac{\rho}{2}|\Delta|\ln\frac{2W}{|\Delta|}\cos\varphi,\nonumber\\
\langle S^y\rangle=0,\\
\langle S^z\rangle=-\rho |\Delta|\frac{g_A}{2g}\cos\varphi.\nonumber
\end{gather}
The finite $\langle S^z\rangle$ also means that a net photocurrent is generated along the edge through the magnetoelectric effect\cite{sajat}. Using the edge Hamiltonian $H=\sum_p 2vpS_p^z$ and introducing a vector potential $A$ to the momentum as an external drive, we can determine by varying $H$ with respect to $A$ that the current density operator is:
\begin{eqnarray}
j(q)=\sum_{p,q}ev(c^{\dagger}_{p+q\uparrow}c_{p\uparrow}-c^{\dagger}_{p+q\downarrow}c_{p\downarrow})e^{-iqr}.
\label{denscur}
\end{eqnarray}
The current operator for $q=0$ is $j=2ev\sum_pS_p^z$, hence:
\begin{eqnarray}
\langle j\rangle=2 ev\langle S^z\rangle=-ev\rho |\Delta|\frac{g_A}{g}\cos\varphi.
\label{curdens}
\end{eqnarray}
The photocurrent is zero when $g_A=0$ which means that the vector potential generates it. This follows from the fact that in Eq.~\eqref{HAM} the vector potential term is similar to an effective magnetic field. Furthermore, we showed earlier that after integrating out the photon field it yields us a vector potential mediated ferromagnetic coupling between spins, which leads to nonzero expectation value for the  magnetization . The direction of the photocurrent is determined by the phase $\varphi$, with $\cos\varphi~=~(-1)^m$.

\section{\label{Pf}Photon Field}

After detailing the effects of the cavity on the material, now we turn to the photon field and how it changes due to the interaction with the edge electrons. To this end, we will study the fluctuations over the mean field parameters which conveniently reveals the validity range of the mean field results\cite{eastham}. We previously presented a physical argument for this, as the effective interactions between electrons have infinite range the mean field results must be exact. Now we make a quantitative argument as well.
Following Ref.[\onlinecite{eastham}] and making use of coherent state path integral formalism we introduce $\phi$ as a complex field for the photons and $\psi_{p\sigma}$ Grassmann fields for the edge electrons. The partition function can be computed as
\begin{eqnarray}
Z=\text{Tr}e^{-\beta H}=\int \mathcal D\phi\int\mathcal D\psi\text{ } e^{-S}
\end{eqnarray}
with action
\begin{eqnarray*}
S=\int_0^\beta d\tau \bar\phi(\partial_\tau+\omega)\phi + \sum_p \bar\eta_p M_p \eta_p = S_{ph}+S_{el}
\end{eqnarray*}
where $\bar\eta_p=\left( \bar\psi_{p\uparrow}\text{ }\bar\psi_{p\downarrow}\right)$ is a spinor and the matrix
\begin{eqnarray*}
M_p=\begin{pmatrix} \partial_\tau + vp+\frac{g_A}{\sqrt L}\text{Re}\phi & \frac{g}{\sqrt L}\phi \\ \frac{g}{\sqrt L}\bar \phi &\partial_\tau - vp-\frac{g_A}{\sqrt L}\text{Re}\phi \end{pmatrix}.
\end{eqnarray*}
Because of superradiance we rescale the photon field $\phi\to\sqrt L \phi$ and integrate out the electron fields. The partition function becomes
\begin{eqnarray}
Z=\int \mathcal D\phi\int\mathcal D\psi\text{ } e^{-S}=\int\mathcal D\phi\text{ }e^{-LS_{eff}} 
\end{eqnarray}
with effective action
\begin{eqnarray}
S_{eff}=\int_0^\beta d\tau \bar\phi(\partial_\tau+\omega)\phi-\frac{1}{L}\ln\left(\int\mathcal D\psi\text{ } e^{-S_{el}}\right).
\end{eqnarray}
If we proceed and try to find the minima of this action ($S_0$) with $\delta S_{eff}/\delta\bar\phi=0$ we arrive at the mean field results Eq.~\eqref{mfp} as $\phi=\sqrt n e^{i\varphi}$. The next step is expanding the effective action around the mean field results to second order which is equivalent to studying the fluctuations around the mean field parameters: $S_{eff}=S_0+S_2(\delta\bar\phi,\delta\phi)$. With this expansion the partition function becomes
\begin{eqnarray}
Z=\int\mathcal D\phi\text{ }e^{-LS_{eff}} =e^{-LS_0}\int\mathcal D\phi\text{ }e^{-LS_2}.
\end{eqnarray}
\jav{Here the $e^{-LS_O}$ term contributes to the mean field result for the free energy. The remaining functional integral gives us the second order correction to the free energy:}
%When calculating the free energy the $e^{-LS_0}$ term will be the mean field result for the free energy and the remaining functional integral gives us the second order correction:
\begin{eqnarray}
F\propto \ln Z=F_{MF}+\frac{1}{L}\ln\det \mathcal D^{-1},
\end{eqnarray}
where $\mathcal D^{-1}$ is the inverse of the Green's function of the photons. It appears because it is the kernel of the action correction $S_2$ and the determinant appears because the functional integral has a simple Gaussian integral form. Since the mean field parameters minimize the effective action this means that $\det\mathcal D^{-1}$ should be positive. In the thermodynamic limit ($L\to\infty$) the correction vanishes thus making the mean field results exact and the superradiant phase as the ground state stable. We will see that $\mathcal D^{-1}$ has zero eigenvalues which describe the Goldstone modes of this system\cite{goldhiggs}, however these modes do not contribute to the free energy in the thermodynamic limit.
 
\subsection{Green's function of the photons}

Instead of calculating the kernel of the second order correction to the effective action, we construct the photon Green's function with diagram technique. Introducing the fluctuations over the mean field parameters we modify Eq.~\eqref{MF} with $a\to\langle a\rangle+a$:
\begin{eqnarray}
\label{PERT}
H=\omega a^\dagger a+\sum_p\begin{pmatrix} c^\dagger_{p\uparrow}&c^\dagger_{p\downarrow}\end{pmatrix}\begin{pmatrix} \varepsilon_p & \Delta\\ \Delta^* &-\varepsilon_p \end{pmatrix}\begin{pmatrix} c_{p\uparrow}\\c_{p\downarrow}\end{pmatrix}\nonumber\\
+\frac{g_A}{\sqrt L}\sum_p \left(a+a^\dagger\right)S_p^z+\frac{g}{\sqrt L}\sum_p \left(aS_p^++a^\dagger S_p^-\right)
\end{eqnarray}
where the first row is the unperturbed mean field Hamiltonian and the second row is understood as the perturbation. In the Nambu space $\left(a\text{ }a^\dagger\right)$ the photon Green's function is
\begin{eqnarray}
\mathcal D(\tau)=-\left\langle T_\tau \begin{pmatrix} a(\tau)a^\dagger(0)&&a^\dagger(\tau)a^\dagger(0)\\
a(\tau)a(0)&&a^\dagger(\tau)a(0)\end{pmatrix}\right\rangle.
\label{ppdef}
\end{eqnarray}
The appearance of anomalous terms are evident from the perturbation as it contains single creation and annihilation photon operators. Because of this, first order diagrams have no contribution and the first non vanishing terms come from second order diagrams, which are single fermion loops. 

%\begin{figure}[h]
%  \centering
%  \begin{tikzpicture}
%    \begin{feynman}
%    \vertex (i1);
% 	\vertex[right=of i1] (a);
%  	\vertex[right=of a] (b);
%  	\vertex[right=of b] (f1);
%      \diagram[horizontal=i1 to f1]{
%      (i1) -- [photon,edge label=$i\omega_n$] (a),
%      (a) -- [fermion, half left, edge label=$i\nu_n+i\omega_n$] (b),
%      (b) -- [fermion, half left, edge label=$i\nu_n$] (a),
%      (b) -- [photon,edge label=$i\omega_n$] (f1),
%      };
%    \end{feynman}
%  \end{tikzpicture}
%  \caption{\label{loop}Single fermion loop with $\omega_n$ photon and $\nu_n$ fermion frequency.}
%\end{figure}
Evaluating these loops in \jav{Matsubara} frequency space using Dyson's equation we arrive at the inverse Green's function for the photons:
\begin{eqnarray*}
\mathcal D^{-1}(i\omega_n)=\mathcal D^{-1}_0(i\omega_n)-\Sigma=\begin{pmatrix}K_1&&K_2\\K_2^*&& K_1^*\end{pmatrix},
\end{eqnarray*}
\begin{eqnarray}
\label{PHG}
K_1=i\omega_n-\omega+\frac{1}{L}\sum_p\Bigg[\frac{4g_A^2|\Delta|^2+g^2(i\omega_n\varepsilon_p +\varepsilon_p^2 +E_p^2)}{E_p(4E_p^2+\omega_n^2)}\nonumber\\-\frac{2gg_A|\Delta|(2\varepsilon_p-i\omega_n)}{E_p(4E_p^2+\omega_n^2)}\Bigg]\tanh\left(\frac{\beta E_p}{2}\right),\nonumber\\
K_2=\frac{1}{L}\sum_p\frac{4g_A^2|\Delta|^2-g^2\Delta^2-2gg_A\Delta\varepsilon_p}{E_p(4E_p^2+\omega_n^2)}\tanh\left(\frac{\beta E_p}{2}\right).\nonumber\\
\end{eqnarray}
Since we are interested in the properties of the ground state of this system we make the $T\to0$ limit and obtain the retarded Green's function as the analytic continuation of Eq.~\eqref{PHG}. 
\subsection{Photon spectral function}\label{subsecB}

The spectral function, defined as the complex part of the trace of the retarded Green's function, is:
\begin{eqnarray}
A(\Omega)=-\frac{1}{\pi}\text{ImTr} \mathcal D(\Omega).
\label{spectral}
\end{eqnarray}
\jav{Carrying out the analytic continuation of Eq.~\eqref{PHG} ($i\omega_n~\to~\Omega~+~i\eta$, with $\eta=0^+$) yields us the following integrals:}
%The analytic continuation of Eq.~\eqref{PHG} means $i\omega_n~\to~\Omega~+~i\eta$, where $\eta=0^+$. The following integrals appear in Eq.~\eqref{PHG}:
\begin{gather}
\frac{1}{L}\sum_p\frac{-4|\Delta|^2}{E_p((\Omega +i\eta)^2-4E_p^2)}=\rho f_0(\Omega),\nonumber\\
\frac{1}{L}\sum_p\frac{\varepsilon_p}{E_p((\Omega +i\eta)^2-4E_p^2)}=0,\label{AC}\\
\frac{1}{L}\sum_p\frac{\varepsilon_p^2}{E_p((\Omega +i\eta)^2-4E_p^2)}=\displaystyle{\frac{\rho}{4} f_2(\Omega)-\frac{\omega}{2g^2}+\frac{\rho g_A^2}{4g^2}}.\nonumber
\end{gather}
The complete forms of $f_0$ and $f_2$ are given in the Appendix. \jav{The properties of these complex valued functions reveal information about the nature of the photon spectral function in Eq.~\eqref{spectral}.} The real parts of \jav{$f_0$ and $f_2$} go to unity when $\Omega$ tends to zero: $\lim_{\Omega\to 0}f_{0,2}(\Omega)=1$. Furthermore, they have vanishing imaginary part when $\Omega<2|\Delta|$ and this sets the threshold energy for continuum polariton excitations for $\Omega>2|\Delta|$.
\jav{Indeed, using the integrals in Eq.~\eqref{AC} the resulting spectral function is zero for frequencies below $2|\Delta|$, except for a well defined $\Omega_0$ value:}
%The spectral function is zero for $\Omega\leq 2|\Delta|$, expect for a well defined $\Omega_0$ value which is understood as the energy of the system's Goldstone mode. Indeed, using the integrals in Eq.~\eqref{AC} the resulting spectral function is: 
\begin{eqnarray}
A(\Omega<2|\Delta|)\propto\text{Im}(F(\Omega)-i\eta)^{-1}=\pi|F'(\Omega_0)|^{-1}\delta(\Omega-\Omega_0).\nonumber
\end{eqnarray}
Here $\Omega_0$ is the real root of the function $F(\Omega)$ defined as: 
\begin{gather}
F(\Omega)=\left(\frac{\rho g_A^2}{2}-\rho g_A^2f_0(\Omega)+\frac{\rho g^2}{2}f_2(\Omega)-\frac{\rho g^2}{4}f_0(\Omega)\right)^2\nonumber\\-\left(\rho g_A^2f_0(\Omega)-\frac{\rho g^2}{4}f_0(\Omega)\right)^2-\Omega^2\left(1-\frac{\rho gg_A}{2|\Delta|}f_0(\Omega)\right)^2\nonumber.\\
\end{gather}
\begin{figure}[t]
\includegraphics[width=6.5cm,height=5cm]{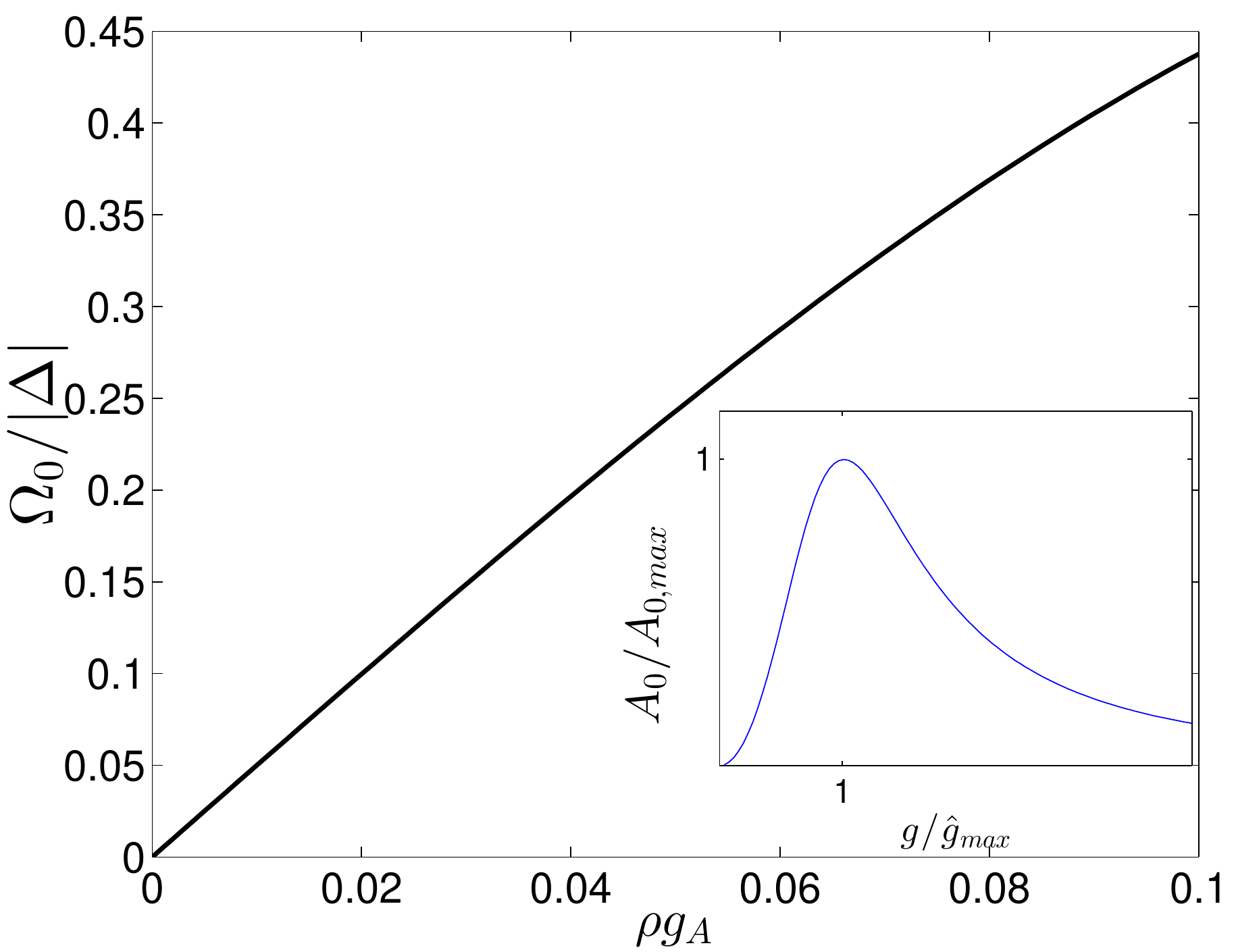}% Here is how to import EPS art
\caption{\label{gold2} The real root of $F(\Omega_0)=0$ as the function of increasing vector potential coupling strength ($g_A$). This is understood as the energy requirement for phase fluctuations and as the energy of the gapped Goldstone mode. The parameters used: $\rho\omega=1$, $\rho g=0.4$. The inset figure is the spectral weight of the Goldstone mode as the function of the Zeeman coupling with parameters: $\rho W=100$, $\rho\omega=1$.}
\end{figure}
Taking $g_A\to0$ in $F(\Omega)$ results in $\Omega_0=0$. When $g_A=0$ the ground state is infinitely degenerate as seen in Fig~\ref{ener}. due to U(1) symmetry and one can sweep through this ground state manifold with no energy cost. This gives rise to a zero energy Goldstone mode which is understood as the phase fluctuation of the superradiant condensate and this appears in the spectral function:
\begin{eqnarray}
A(\Omega<2|\Delta|)=\frac{6\rho g^2|\Delta|^2}{12|\Delta|^2 +(\rho g^2)^2}\delta '(\Omega)=A_0\delta'(\Omega).
\end{eqnarray}
%For very small $g_A\ll g$ the spectral function in Eq.~\ref{spectral}. has the form: $A(\Omega<2|\Delta|)=$
%\begin{eqnarray}
%\frac{3|\Delta |^2(\rho g^2-2\rho g_A^2)\delta(\Omega-\Omega_0)}{\Omega_0\left(\rho^2 g_A^2 g^2+2(\rho %g_A^2)^2+(\rho g^2)^2+12|\Delta |^2-12\rho gg_A|\Delta | \right)},\nonumber\\
%\end{eqnarray}

The spectral weight of the Goldstone mode vanishes with $g$. Since the gap depends on $g$ according to Eq.~\eqref{mfp} it has a maxima ($A_{0,max}$) at the solution of $\ln\frac{48W^2}{x(x-\omega)}=\frac{4\omega}{x}$ for $x=\rho \hat g_{max}^2$ and vanishes as $g$ increases. This is shown in the inset of Fig.~\ref{gold2}. In the presence of a nonzero $g_A$ phase fluctuations will require finite amount of energy, thus making the Goldstone mode gapped which is described in $\Omega_0$, see Fig~\ref{gold2}. 
\begin{figure}[t]
\includegraphics[width=6.5cm,height=5cm]{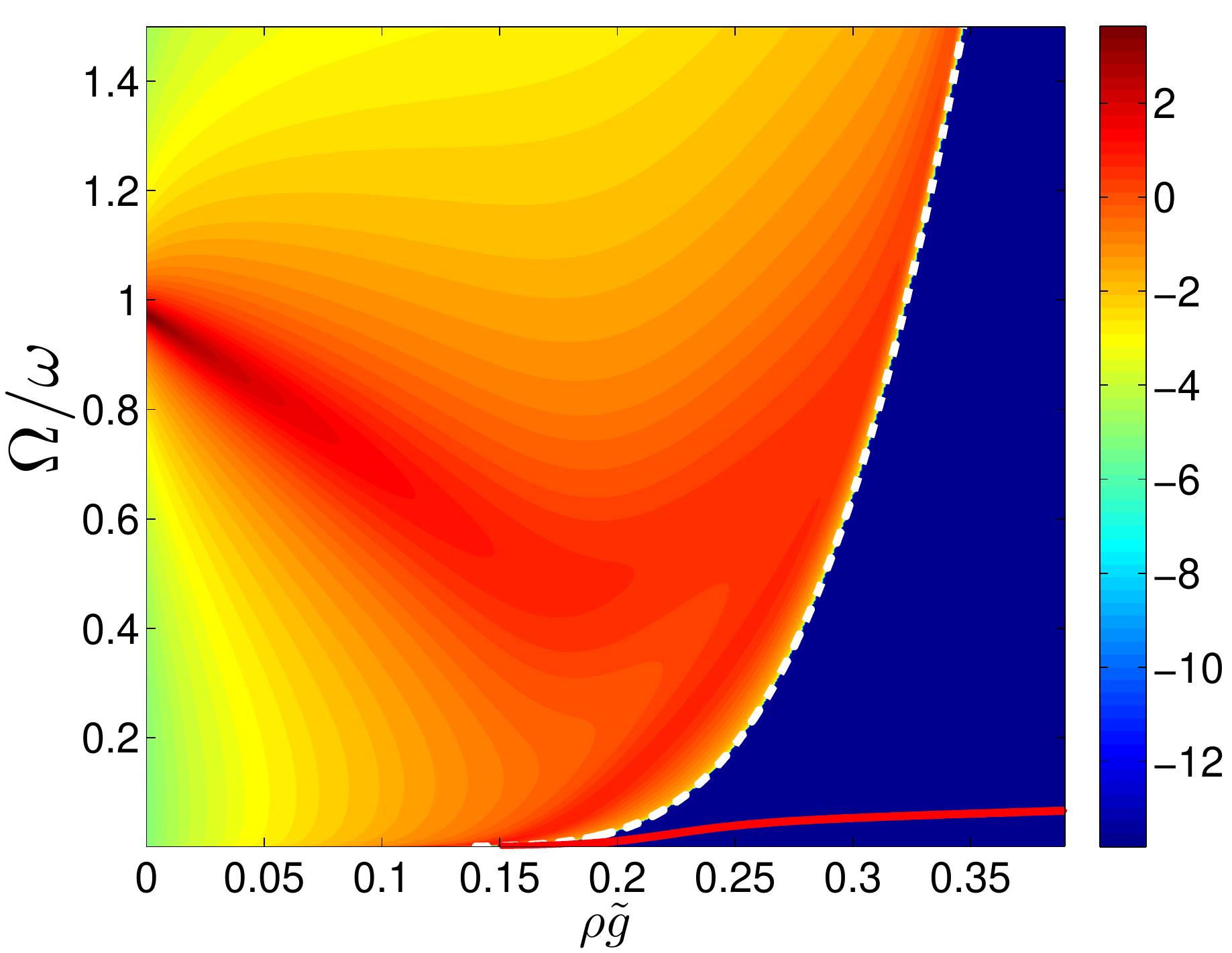}% Here is how to import EPS art
\caption{\label{spect} The contour plot of the spectral function $\ln(\omega A(\Omega))$ as the function of the Zeeman coupling $\rho\tilde g$ with $g_A=0.1$. The energy of the Goldstone mode $\Omega_0$ is shown as the red solid line in the $\Omega<2|\Delta|$ regime. The white dashed line denotes the minimum excitation energy $2|\Delta|$, above which the polariton continuum is formed. The parameters used: $\rho W=100$, $\rho\omega=1$.}
\end{figure}

In the $\Omega>2|\Delta|$ case the complex part of the functions $f_0$ and $f_2$ are nonzero and we get the polariton excitations and their spectral weight in the spectral function $A(\Omega>2|\Delta|)$, which is measurable by the absorption coefficient of the cavity.
Without interactions the spectral function has the form $A(\Omega)=\delta(\Omega-\omega)$ with the bare $\omega$ cavity mode. On Fig.~\ref{spect}. it is noticable that this mode is shifted down from $\omega$ because of $g_A$ and with increasing $g$ it gets damped as $\sim\rho g^2$. Eventually it renormalizes into smaller frequencies before hitting the optical gap at $\Omega=2|\Delta|$, where the spectral function exhibits a square root singularity. Apart from shifting $\omega$ for small $g$ the vector potential coupling does not have significant contribution to the nature of the polariton continuum.

\section{Conductivity along the edge}

Equipped with the photon Green's function, we can evaluate the Kubo formula for the frequency dependent \jav{optical conductivity along the edge. The density-density correlation function, which is readily related  to  the  optical  conductivity, can  be  investigated  by shot  noise  measurements. In addition, the optical conductivity can directly be probed by the amplitude or phase modulation of the optical lattice\cite{optcond,optcond1}, that realizes the spin Hall insulator of our system.} 

The response for an external drive have two contributions: 
\begin{eqnarray}
\sigma=\sigma_{Kubo}+\sigma_{dia}.
\label{fullcond}
\end{eqnarray}
The first is the direct result for the conductivity computed from the Kubo formula:
\begin{eqnarray}
\sigma_{Kubo}=\frac{\chi(\Omega)}{i\Omega},
\end{eqnarray}
where $\chi$ is the current-current correlation function.
\begin{eqnarray}
\chi(\tau)=\langle T_{\tau}j(\tau)j(0)\rangle =4e^2v^2 \sum_p\langle T_\tau S_p^z(\tau)S_p^z(0)\rangle.
\label{corr}
\end{eqnarray}
The second term in Eq.~\eqref{fullcond} is a diamagnetic term. By diagonalising the Hamiltonian in the presence of an external vector potential, the resulting spinor wavefunctions will depend on the vector potential through the Peierls substitution. 
Calculating the expectation value of the current operator to first order in the vector potential gives us the diamagnetic contribution in the conductivity formula: $\sigma_{dia}=-\rho e^2v^2/(i\Omega)$. This is akin to the origin of a diamagnetic 
term in graphene where the energy dispersion is also linear\cite{geimgraf,condgraf}.

We calculate the correlation function in Eq.~\eqref{corr} diagramatically in Matsubara frequency space. The diagrams we need to consider are a single fermion loop and a collective mode diagram\cite{lee}.
%in Fig.~\ref{currdiag}.
%\begin{figure}[h]
%  \centering
%  \begin{tikzpicture}
%    \begin{feynman}
%    \vertex (i1);
%    \vertex[above=0.5cm of i1]{a.,};
% 	\vertex[right=1cm of i1](f1);
% 	\vertex[right=1cm of f1](x);
%      \diagram[horizontal=i1 to f1]{
%      	(i1) --[fermion, half left] (f1),
%      	(f1) --[fermion, half left] (i1),
%      	(f1) --[white] (x)
%      };
%    \end{feynman}
% \end{tikzpicture}
% \begin{tikzpicture}
%  	\begin{feynman}
%	\vertex (i1);
%	\vertex[above=0.5cm of i1]{b.,};
% 	\vertex[right=1cm of i1](f1);
% 	\vertex[right=1cm of f1](i2);
% 	\vertex[right=1cm of i2](f2);
%      \diagram[horizontal=i1 to f2]{
%      	(i1) --[fermion, half left] (f1),
%      	(f1) --[fermion, half left] (i1),
%      	(f1) --[photon] (i2),
%      	(i2) --[fermion, half left] (f2),
%      	(f2) --[fermion, half left] (i2),
%      };
%	\end{feynman}
%	\end{tikzpicture}
%  \caption{\label{currdiag}a., Single particle contribution and b., the collective contribution to the current-current correlation function.}
%\end{figure}
%The first diagram is a simple bubble, evaluating this we get 
Evaluation of the single fermion loop gives us:
\begin{eqnarray}
\Pi_{zz}^0(i\omega_n)=-\frac{T}{L}\sum_{p,\nu_n}\text{Tr}[\sigma^z\mathcal{G}(i\nu_n)\sigma^z \mathcal{G}(i\nu_m-i\omega_n)].\text{ }
\label{pi0}
\end{eqnarray} 
Here $\mathcal{G}$ is the electron Green's function, using Eq.~\eqref{bog} and Eq.~\eqref{HMF} this reads as:
\begin{eqnarray}
\mathcal{G}(p,i\nu_n)=\frac{1}{(i\nu_n)^2-E_p^2}\begin{pmatrix}
i\nu_n +\varepsilon_p && \Delta\\
\Delta ^* && i\nu_n -\varepsilon_p
\end{pmatrix}.
\end{eqnarray}
The term $\Pi_{zz}$ is related to the correlation function in Eq.~\eqref{corr} as $\chi=v^2\Pi$. Summing over the frequencies and momenta in Eq.~\eqref{pi0} and taking the temperature to zero, we get:
\begin{eqnarray}
\Pi_{zz}^0(\Omega)=\rho f_0(\Omega).
\end{eqnarray}
To evaluate the collective diagram we need to construct the RPA equations. 
Instead of using $\mathcal{D}_0$ the unperturbed photon propagator and consider a connected RPA system of equations, we follow here a different approach. 
Since we already calculated the full photon propagator in Eq.~\eqref{PHG}, we sum up all the possible combinations that would appear from the interaction term of Eq.~\eqref{PERT}. This immediately gives us the correlation function:
\begin{widetext}
\begin{gather}
\Pi_{zz}=\Pi_{zz}^0-g^2\left(\Pi_{z+}^0\mathcal D_{aa}\Pi_{+z}^0+\Pi_{z+}^0\mathcal D_{aa^\dagger}\Pi_{-z}^0+\Pi_{z-}^0\mathcal D_{a^\dagger a}\Pi_{+z}^0+\Pi_{z-}^0\mathcal D_{a^\dagger a^\dagger}\Pi_{-z}^0\right)\nonumber\\
-g_A^2\left(\Pi_{zz}^0\mathcal D_{aa}\Pi_{zz}^0+\Pi_{zz}^0\mathcal D_{aa^\dagger}\Pi_{zz}^0+\Pi_{zz}^0\mathcal D_{a^\dagger a}\Pi_{zz}^0+\Pi_{zz}^0\mathcal D_{a^\dagger a^\dagger}\Pi_{zz}^0\right)\nonumber\\
-gg_A\left(\Pi_{z+}^0\mathcal D_{aa}\Pi_{zz}^0+\Pi_{z+}^0\mathcal D_{aa^\dagger}\Pi_{zz}^0+\Pi_{z-}^0\mathcal D_{a^\dagger a}\Pi_{zz}^0+\Pi_{z-}^0\mathcal D_{a^\dagger a^\dagger}\Pi_{zz}^0\right)\nonumber\\
-g_Ag\left(\Pi_{zz}^0\mathcal D_{aa}\Pi_{+z}^0+\Pi_{zz}^0\mathcal D_{aa^\dagger}\Pi_{-z}^0+\Pi_{zz}^0\mathcal D_{a^\dagger a}\Pi_{+z}^0+\Pi_{zz}^0\mathcal D_{a^\dagger a^\dagger}\Pi_{-z}^0\right).
\label{pizza}
\end{gather}
\end{widetext}
The minus signs in front of the couplings come from the definition of the photon propagator in Eq.~\eqref{ppdef}. In Eq.~\eqref{pizza} there are four more frequency sums:
\begin{eqnarray}
%\Pi_{z\pm}^0(i\omega_n)=-\frac{T}{L}\sum_{p,\nu_n}\text{Tr}[\sigma^z\mathcal{G}(i\nu_n)\sigma^\pm \mathcal{G}(i\nu_m-i\omega_n)],\nonumber\\
\Pi_{ab}^0(i\omega_n)=-\frac{T}{L}\sum_{p,\nu_n}\text{Tr}[\sigma^a\mathcal{G}(i\nu_n)\sigma^b \mathcal{G}(i\nu_m-i\omega_n)],\nonumber
\end{eqnarray} 
with $a$ and $b$ are $z$ or $\pm$. Doing the same procedure as in Eq.~\eqref{pi0} these cross correlations are:
\begin{eqnarray}
\Pi_{\pm z}^0(\Omega)=\mp\Pi_{z\mp}^0(\Omega)=\pm\frac{\rho\Omega}{4|\Delta|}f_0(\Omega).\nonumber
\end{eqnarray}
To summarize Eq.~\eqref{pizza} we gather every term into a single function:
\begin{eqnarray}
\Pi_{zz}(\Omega)=\rho \left[f_0(\Omega)-C(\Omega)f_0(\Omega)\right],
\end{eqnarray}
and we arrive at the full \jav{optical conductivity formula}:
\begin{eqnarray}
\sigma(\Omega)=\frac{\rho e^2 v^2}{i\Omega}\left[f_0(\Omega)-1-C(\Omega)f_0(\Omega)\right].
\end{eqnarray} 
This expression is very similar to other conductivity formulas for electron-phonon coupled systems calculated with RPA\cite{virobacsi,vanyolos2}. 

\begin{figure}[t]
\includegraphics[width=6.5cm,height=5cm]{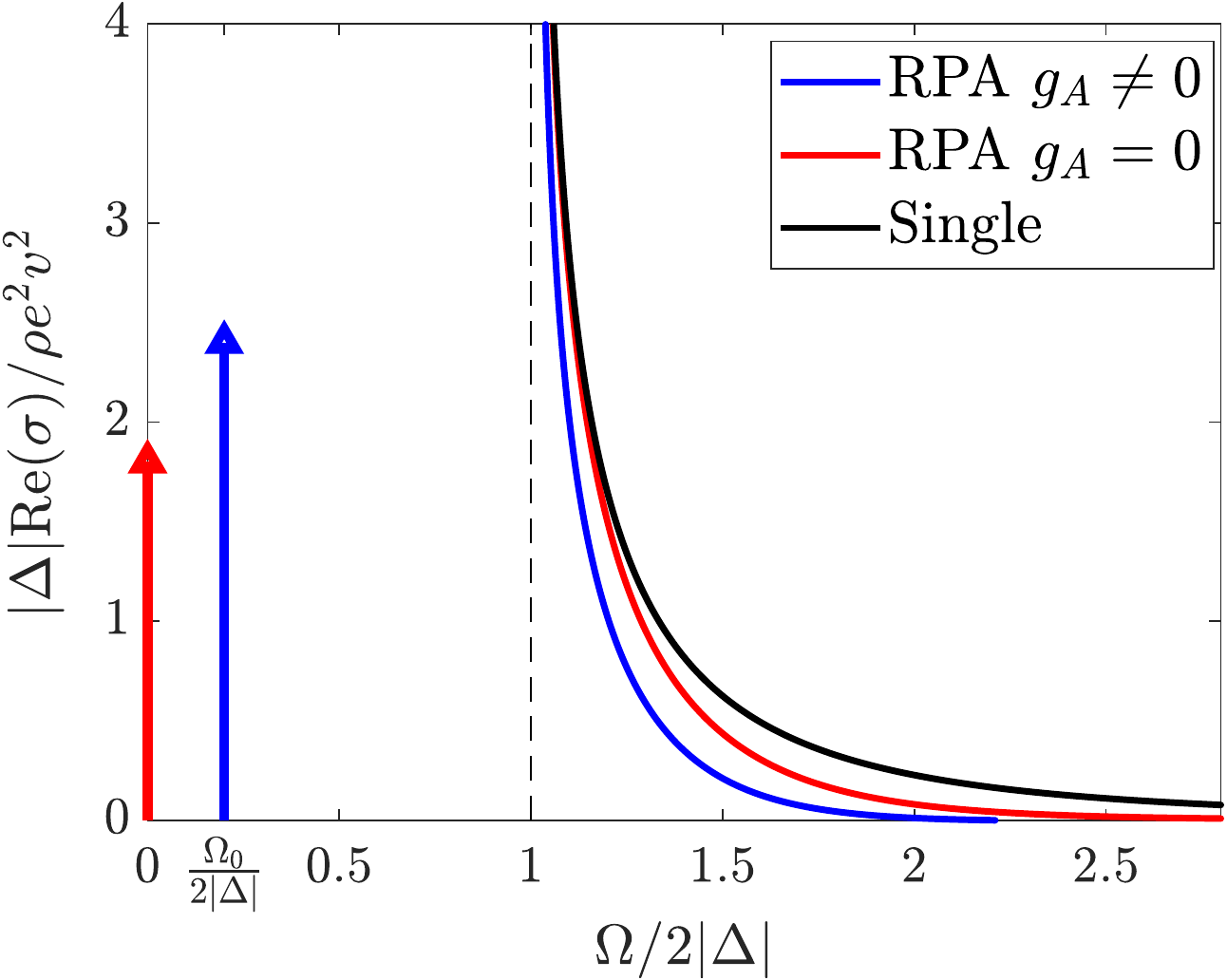}% Here is how to import EPS art
\caption{\label{cond} The frequency dependent conductivity of the edges. The solid black is the single particle result, the red contains the collective terms with $\rho g_A=0$ and the blue contains all terms with $\rho g_A=0.1$. The Goldstone peaks are also depicted, as $g_A$ is nonzero the peak moves to frequency $\Omega_0$ and its weight increases. Further parameters: $\rho\omega=1,\rho W=100, \rho g=0.4$.}
\end{figure}

Let us first examine the properties of the conductivity through the function $C$ when $g_A=0$. In this case we need to condsider the first row of Eq.~\eqref{pizza}, the function $C$ has the form:
\begin{eqnarray}
C(\Omega)=-\frac{\Omega^2}{4|\Delta|^2}\frac{f_0(\Omega)f_2(\Omega)}{f_2(\Omega)(f_2(\Omega)-f_0(\Omega))-\frac{4}{\rho^2g^4}\Omega^2 -i\eta}.\nonumber
\end{eqnarray}
When $\Omega\to 0$, $C(0)=\rho^2g^4/(\rho^2g^4+16|\Delta|^2)$. By the Kramers--Kronig relation, this implies a Dirac delta function at the origin of the real part of the conductivity. 
Indeed making the $\eta\to0^+$ limit we get the Dirac delta in accordance with Kramers--Kronig. This result clearly comes from the full photon propagator and is absent from the single particle contribution to the optical response, therefore 
the Goldstone mode manifests itself in the conductivity formula as a Drude peak:
\begin{eqnarray}
\sigma_{Goldstone}=\pi\rho e^2v^2\frac{\rho^2 g^4}{\rho^2g^4+16|\Delta|^2}\delta(\Omega).
\end{eqnarray}
If we take the $g\to0$ then $\sigma_{Goldstone}=\pi\rho e^2v^2\delta(\Omega)$, so it becomes the conventional Drude weight\cite{drude}. 
This allows us to introduce an effective mass due to light-matter interaction. The Drude weight of the non-interacting system reads as $n_ee^2/m$,
 where $n_e$ is the particle number density of the edge electrons
 and $m$ is their mass. In the presence of interaction, we rewrite the Goldstone conductivity  as
\begin{eqnarray}
\sigma_{Goldstone}=\frac{\pi n_e e^2}{m^*}\delta(\Omega),
\end{eqnarray}
with effective mass:
\begin{eqnarray}
\frac{m}{m^*}=\frac{\rho^2 g^4}{\rho^2g^4+16|\Delta|^2}.
\label{mecs}
\end{eqnarray}
 The $C$ function is a combination of the previously defined $f_{0,2}$ functions, which indicates that the real part of the conductivity must be zero for frequencies below $2|\Delta|$. The behaviour of Re$(\sigma)$ is shown in Fig.~\ref{cond}., with $C=0$ the single particle term has a square root singularity at frequency twice the gap. Considering the collective modes the square root singularity still remains, however a portion of the weight of the conductivity is transferred into the weight of the Goldstone mode, so that the conductivity sum rule is not violated, indeed:
\begin{eqnarray}
\int_{0}^\infty \text{d}\Omega \text{ Re}(\sigma(\Omega))=\frac{\pi}{2}\rho e^2v^2.
\label{sum}
\end{eqnarray}
Turning now to the case when $g_A$ is nonzero the function $C$ is given by Eq.~\eqref{pizza}. Notice that in Eq.~\eqref{corr} for convenience we used a time ordered product instead of a commutator in the Kubo 
formula. Unless the current operator possess a nonzero expectation value\cite{electronliquid}, these two approaches give the same result. However, in Eq.~\eqref{curdens}, the current operator has a finite expectation value in the ground-state, which means 
that our result contains an extra term in Eq.~\eqref{pizza}, which is only present in the time ordered product but should be absent from the commutator:
\begin{eqnarray}
-\frac{\rho^2 g_A^4 f_0(\Omega)}{F(\Omega)-i\eta}.
\end{eqnarray}
This we must neglect\cite{electronliquid}. The correct expression, in accordance with the linear response commutator from the Kubo formula, is:
\begin{gather}
C(g_A,\Omega)=-\frac{\rho f_0(\Omega)}{F(\Omega)-i\eta}\Big[\frac{g^2\Omega^2}{16|\Delta|^2}\big(\rho g_A^2-4\rho g_A^2f_0(\Omega)\nonumber\\ +\rho g^2 f_2(\Omega)\big)
+\rho g_A^2 g^2(f_2(\Omega)-f_0(\Omega))\nonumber \\ +\frac{gg_A\Omega^2}{|\Delta|}\left(1-\frac{\rho gg_A}{2|\Delta|}f_0(\Omega)\right)\Big].
\label{cga}
\end{gather}
When $\Omega\to0$ Eq.~\eqref{cga} vanishes and thus the Drude peak disappears. However, the real part of the conductivity still has a Dirac delta at the frequency where $F(\Omega)=0$, which corresponds to the gapped Goldstone mode energy $\Omega_0$:
\begin{eqnarray}
\sigma_{Goldstone}=\frac{\pi n_e e^2}{m^*}\delta(\Omega-\Omega_0),
\end{eqnarray}
with effective mass that depends on the energy of the gapped Goldstone mode:
\begin{gather}
\frac{m}{m^*}=\frac{\Omega_0 f_0(\Omega_0)^2}{16|\Delta|^2|F'(\Omega_0)|}\bigg[\rho^2 g^4 f_2(\Omega_0)+16\rho gg_A|\Delta|\nonumber\\
+\rho^2g^2g_A^2\big(1-16f_0(\Omega_0))\big)\bigg].
\label{veff}
\end{gather}
\begin{figure}[t]
\includegraphics[width=6.5cm,height=5cm]{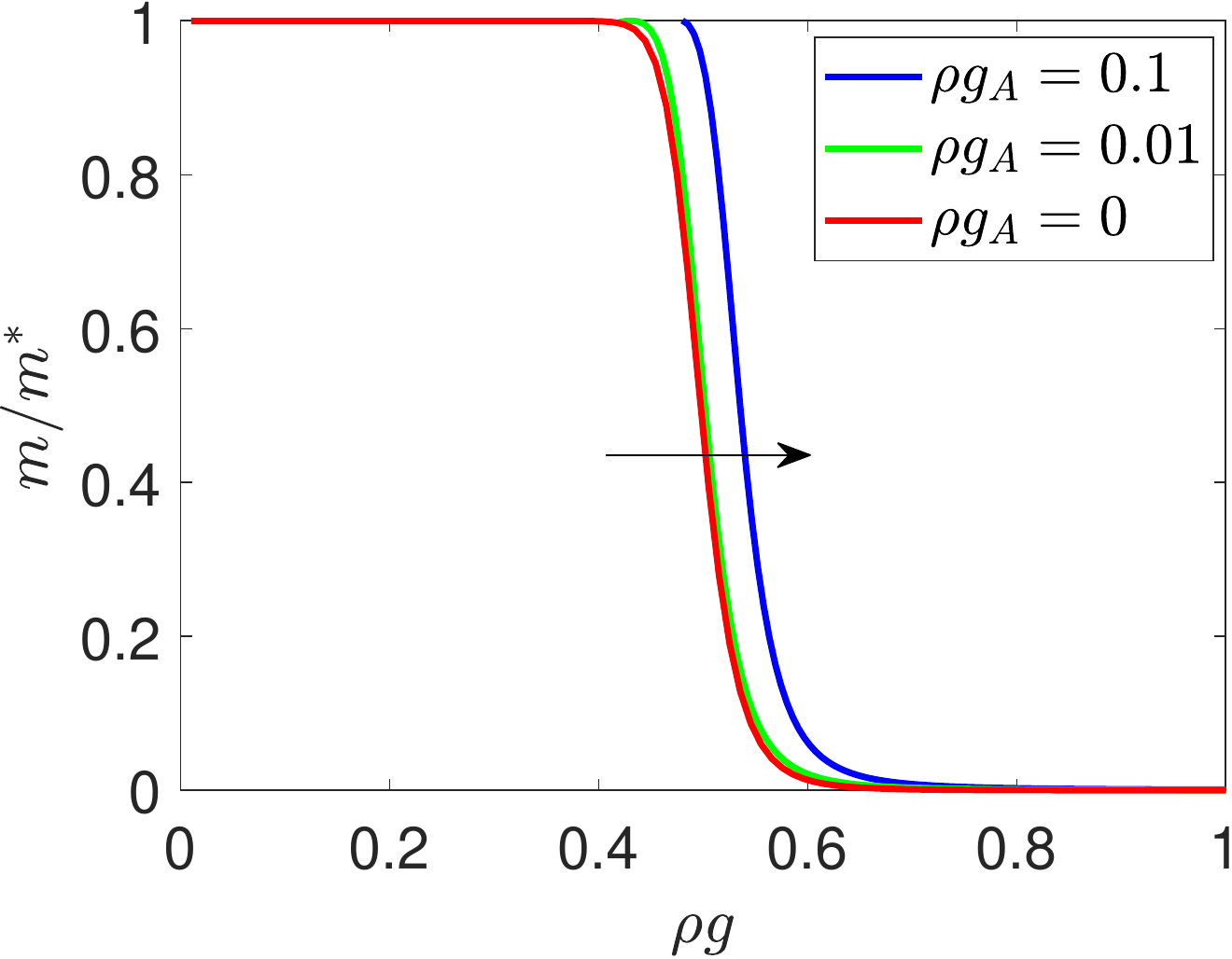}% Here is how to import EPS art
\caption{\label{meff} The effective mass as the function of the Zeeman interaction strength is plotted and it resembles a step function. The arrow indicates that the increase of $g_A$ shifts this step, thus the effective mass becomes infinite at larger Zeeman coupling strength. This means that the interaction involving the vector potential is making the collective modes more stable at larger $g$. The parameters used: $\rho\omega=1,\rho W=100.$}
\end{figure}
Instead of a dc conductivity we get a low frequency ac one at $\Omega_0$. These results are very similar to the interband conductivity obtained when studying electron interaction with Fr\"ohlich phonons, there the resulting dc
 conductivity becomes a low frequency ac due to Coulomb interactions\cite{lee}.

In the absence of interactions, the real part of the conductivity of the edge \jav{electrons} consists of only the bare Drude peak with mass $m$. As the Zeeman interaction appears the weight of this Drude peak decreases ($m^*$ increases) and the real part of the conductivity is now nonzero for frequencies over $2|\Delta|$. As $g$ grows so does the effective mass and when the coupling strength $g$ is comparable with the photon frequency ($g\approx\omega$) the effective mass renormalizes to nearly infinity, see Fig.~\ref{meff}, thus making the collective modes in the conductivity disappear. This means that the single particle description of the conductivity is sufficient in this parameter range. In addition to shifting the Drude peak to frequency $\Omega_0$, the appearance of $g_A$ also decreases the effective mass $m^*$. This can be seen on Fig.~\ref{cond}., as the conductivity curve when $g_A$ is nonzero is always under the curve of the zero $g_A$ case. The missing weight is transfered into the the weight of the Goldstone mode, due to the conductivity sum rule in Eq.~\eqref{sum}, $m^*$ must decrease. This means that the vector potential interaction stabilizes the collective modes at stronger Zeeman couplings. Fig.~\ref{meff}. also supports this idea. 

\section{Conclusion}

Interaction between a circularly polarized quantum photon field and spin Hall edge \jav{electrons} leads to a stable superradiant ground state at arbitrary Zeeman interaction strength. 
This ground state spontanously breaks time reversal symmetry 
and a net photocurrent or equivalently magnetization along axis-z through the magnetoelectric coupling, is generated by the vector potential part 
of the electromagnetic field. Above a threshold energy, corresponding to a Higgs mode, a continuum polariton excitations emerge from the single cavity mode and below the threshold a 
Goldstone mode arises from the phase fluctuations of the ground state. Without the coupling to the vector potential this mode sits at zero energy  due to the broken continuous U(1) symmetry. 
The introduction of the vector potential decreases the symmetry of the system into discrete time reversal. 
This results in a gapped Goldstone mode as phase fluctuations require a finite amount of energy to connect the symmetry broken ground states. 
In an external classical electromagnetic field, this Goldstone mode manifests itself in the frequency dependent conductivity along the edges and produces a low frequency dc/ac conductivity, depending on the absence/presence of 
the vector potential term, respectively. 
When the Zeeman coupling becomes comparable with the photon frequency, these conductivity structures  only survive if the interaction involving the vector potential is present. 
For larger frequencies, the conductivity is zero for frequencies smaller than twice the gap, and has  a characteristic square root singularity at the Higgs mode, $\Omega=2|\Delta|$ and vanishes for increasing frequencies.
\javi{Finally, we remark that the requirement for the observation of the superradiant phase that the temperature should be well below the gap size. Similarly to other predictions made by mean field theory, the transition temperature is always comparable to the gap size\cite{bruus} and as such for temperatures $T\ll\Delta$ the effects detailed above should be observable.}
 
\begin{acknowledgments}
This research is supported by the National Research, Development and Innovation Office - NKFIH within the Quantum Technology National Excellence Program (Project No.
      2017-1.2.1-NKP-2017-00001), K119442, by the  BME-Nanonotechnology FIKP grant of EMMI (BME FIKP-NAT) and by Romanian UEFISCDI, project number PN-III-P4-ID-PCE-2016-0032.
\end{acknowledgments}

\section{Appendix}
\jav{
\subsection{Derivation of the Hamiltonian in Eq.~\eqref{HAM}}
}
%\jav{
%Our system involves spin filtered edge electrons with linear momentum: $\varepsilon_\sigma(p)=\sigma vp $. With $c^\dagger_{p\sigma}$ creating an edge electron with momentum $p$ and $\sigma=\uparrow,\downarrow$ spin:
%\begin{gather}
%H_{edge}=\sum_{p,\sigma} \varepsilon_\sigma(p)c_{p\sigma}^\dagger c_{p\sigma}=\sum_p vp(c_{p\uparrow}^\dagger c_{p\uparrow}-c_{p\downarrow}^\dagger c_{p\downarrow})
%\end{gather}
%}
%\jav{The edge electrons are placed inside a cavity, that having its own quantum dynamics. The quantized vector potential is:
%\begin{gather}
%A(r)=\sum_{k,\mu} (a_{k\mu} e^{ikr}e_\mu+a_{k\mu}^\dagger e^{-ikr}e_\mu^*)
%\end{gather}
%}
%\jav{Here $a^\dagger_{k\mu}$ creates a photon with momentum $k$ and helicity $\mu$. We are interested in the interaction between edge electrons and a single mode of electromagnetic field with fixed helicity. The interaction originates from a Zeeman coupling between electron spins and the magnetic field:
%\begin{gather}
%H_{int;1} =\sum_{\alpha,\beta}\int \mathrm{d}^3r\,\Psi^\dagger_\alpha(\mathbf r)\left(g_e\mu_B\mathbf S\cdot\mathbf B(\mathbf r)\right)\Psi_\beta(\mathbf r)\nonumber\\=\frac{g}{\sqrt L}\sum_p\left(aS_p^++a^\dagger S_p^-\right).
%\end{gather}
%}
\jav{
Our system involves spin-momentum locked edge electrons with linear momentum: $\varepsilon_\sigma(p)=\sigma vp,\, \sigma=\pm 1$ and $v$ is the Fermi velocity. With the definition of $c^\dagger_{p,\sigma}$ which creates an edge electron with  momentum~$p$ and spin~$\sigma$, we have:
\begin{gather}
H_{edge}=\sum_{p,\sigma} \varepsilon_\sigma(p)c_{p\sigma}^\dagger c_{p\sigma}=\sum_p vp(c_{p\uparrow}^\dagger c_{p\uparrow}-c_{p\downarrow}^\dagger c_{p\downarrow}).
\end{gather}}
\jav{The edge electrons are placed inside a cavity (Fig.~\ref{rajz}), that is having its own quantum dynamics. We are interested in the interaction of the edge electrons and a single mode of quantum light with fixed helicity. The energy of the mode is: $H_{field}=\omega a^\dagger a$, where $\omega$ and $a$ denote the frequency and annihilation operator of a photon with positive helicity, respectively.} 
\jav{The interaction arises from the magnetic part of the electromagnetic field that interacts with the spin of an edge electron. This is a Zeeman interaction:
\begin{gather}
\label{zeeman}
H_{Z} =\sum_{\alpha,\beta}\int \mathrm{d}^3r\,\Psi^\dagger_\alpha(\mathbf r)\left(g_e\mu_B\mathbf S\cdot\mathbf B(\mathbf r)\right)\Psi_\beta(\mathbf r)\nonumber\\=\frac{g}{\sqrt L}\sum_p\left(aS_p^++a^\dagger S_p^-\right).
\end{gather}
}
\jav{Here $g=\frac{g_e\mu_B}{2c}\sqrt{\frac{\omega}{\varepsilon_0}}\equiv\sqrt{\tilde g \omega}$ is the coupling constant of the Zeeman term, we used: 
\begin{gather}
\Psi_\sigma(\mathbf r)=\frac{1}{\sqrt L}\sum_p c_{p\sigma}e^{ipr}, \nonumber\\  \mathbf B(\mathbf r)=\frac{i}{c}\sqrt{\frac{\omega}{2\varepsilon_0 L}}\left(\left(\mathbf e_z\times\mathbf e_{+}\right)ae^{ikr}-\left(\mathbf e_z\times\mathbf e_{+}^*\right)a^\dagger e^{-ikr}\right)\nonumber\\
\mathbf S=\frac{1}{2}(\sigma^x\,\sigma^y\,\sigma^z)^T,\,\, S_p^\pm=\frac{1}{2}\sum_{\alpha,\beta}c^\dagger_{p,\alpha}\left(\sigma^x_{\alpha,\beta}\pm i\sigma^y_{\alpha,\beta}\right)c_{p,\beta}.
\end{gather}}
\jav{There is another interaction term present from the vector potential of the quantum electromagnetic field due to the Peierls substitution $(p\to p+eA)$:
\begin{gather}
\label{vect}
H_{A}= \sum_{\alpha,\beta}\int \mathrm{d}^3r\,\Psi^\dagger_\alpha(\mathbf r)\left(ev\sigma_{\alpha,\beta}^z A_x(\mathbf r)\right)\Psi_\beta(\mathbf r)\nonumber\\=\frac{g_A}{\sqrt L}\sum_p\left(a+a^\dagger\right)S_p^z,
\end{gather}
}
\jav{The final Hamiltonian in Eq.~\eqref{HAM} is therefore:
\begin{gather}
 H=H_{field}+H_{edge}+H_{A}+H_{Z}
\end{gather}}
\subsection{The complete forms of the functions $f_{0;2}$ in Eq.~\eqref{AC}}
The functions $f_0$ and $f_2$ can be calculated from Eq.~\eqref{AC} with Eq.~\eqref{Ep}: 
\begin{gather}
f_0(\Omega)=\frac{1}{\rho L}\sum_p\frac{-4|\Delta|^2}{E_p((\Omega +i\eta)^2-4E_p^2)}=\nonumber \\
-2v|\Delta|^2\int_{-W}^W\text{d}p\frac{1}{E_p((\Omega +i\eta)^2-4E_p^2)}
\end{gather}
and

\begin{gather}
f_2(\Omega)=\frac{2\omega}{\rho g^2}+\frac{g_A^2}{g^2}+\frac{1}{\rho L}\sum_p\frac{4\varepsilon_p^2}{E_p((\Omega +i\eta)^2-4E_p^2)}=\nonumber \\
\frac{2\omega}{\rho g^2}+\frac{g_A^2}{g^2}+2v\int_{-W}^W\text{d}p\frac{\varepsilon_p^2}{E_p((\Omega +i\eta)^2-4E_p^2)}.
\end{gather}
By carrying out the integration with respect to the momentum $p$ and disregard terms that is the order or lower than $W^{-1}$, we get the complete forms of $f_0$ and $f_2$ with $\Theta(x)$ Heaviside functions: 
\begin{widetext}
\begin{eqnarray}
f_0(\Omega)=\frac{\Theta(2|\Delta|-\Omega)4|\Delta|^2}{\Omega\sqrt{4|\Delta|^2-\Omega^2}}\text{arctg}\frac{\Omega}{\sqrt{4|\Delta|^2-\Omega^2}}+\Theta(\Omega-2|\Delta|)\left[\frac{2i\pi|\Delta|^2}{\Omega\sqrt{\Omega^2-4|\Delta|^2}}-\frac{4|\Delta|^2}{\Omega\sqrt{\Omega^2-4|\Delta|^2}}\text{arth}\frac{\sqrt{\Omega^2-4|\Delta|^2}}{\Omega}\right]\nonumber\\
f_2(\Omega)=\Theta(2|\Delta|-\Omega)\frac{\sqrt{4|\Delta|^2-\Omega^2}}{\Omega}\text{arctg}\frac{\Omega}{\sqrt{4|\Delta|^2-\Omega^2}}+\Theta(\Omega-2|\Delta|)\left[\frac{\sqrt{\Omega^2-4|\Delta|^2}}{\Omega}\text{arth}\frac{\sqrt{\Omega^2-4|\Delta|^2}}{\Omega}-\frac{i\pi\sqrt{\Omega^2-4|\Delta|^2}}{2\Omega}\right]\nonumber
 \label{eq:wideeq}
\end{eqnarray}
\end{widetext}

\bibliography{regraph.bib}

\end{document}